\documentclass[10pt, twocolumn]{article}   	
\usepackage{geometry}                		
\pdfoutput=1                 		
\usepackage{graphicx}
\usepackage{subfigure}				
\usepackage{amssymb}
\usepackage{amsmath}
\usepackage{url}
\usepackage{hyperref}
\title{\bf Time Series Vector Autoregression Prediction of the Ecological Footprint based on Energy Parameters}
\author{Radmila Jankovi\'c$^1$, Ivan Mihajlovi\'c$^2$, Alessia Amelio$^3$}
\date{\small $^1$Mathematical Institute of the S.A.S.A, Belgrade, Serbia, \href{mailto:rjankovic@mi.sanu.ac.rs}{\underline{rjankovic@mi.sanu.ac.rs}}\\$^2$Technical Faculty in Bor, University of Belgrade, Bor, Serbia, \href{mailto:imihajlovic@tfbor.bg.ac.rs}{\underline{imihajlovic@tfbor.bg.ac.rs}}\\$^3$DIMES, University of Calabria, Rende (CS), Italy, \href{mailto:aamelio@dimes.unical.it}{\underline{aamelio@dimes.unical.it}}	}						

\begin{document}
\maketitle

{\small \bf \textit{Abstract} -- Sustainability became the most important component of world development, as countries worldwide fight the battle against the climate change. To understand the effects of climate change, the ecological footprint, along with the biocapacity should be observed. The big part of the ecological footprint, the carbon footprint, is most directly associated with the energy, and specifically fuel sources. This paper develops a time series vector autoregression prediction model of the ecological footprint based on energy parameters. The objective of the paper is to forecast the EF based solely on energy parameters and determine the relationship between the energy and the EF.  The dataset included global yearly observations of the variables for the period 1971-2014. Predictions were generated for every variable that was used in the model for the period 2015-2024. The results indicate that the ecological footprint of consumption will continue increasing, as well as the primary energy consumption from different sources. However, the energy consumption from coal sources is predicted to have a declining trend.}

\section{Introduction}
According to the OECD (The Organisation for Economic Co-operation and Development) report \cite{[1]}, the industrial production and construction involving mining, manufacturing, electricity, gas, and air-conditioning sectors in 2018 had an increasing trend, compared to 2015. Moreover, in the first quarter of 2019, global manufacturing increased by 2.5 percent compared to the first quarter of 2018 \cite{[2]}. At the same time, in 2018, global primary energy consumption grew by 2.9 percent comparing to the year before, which was the fastest observed growth since 2010 \cite{[3]}. In terms of the fuel type, all fuels had a highly increasing trend compared to their 10-year averages, except for renewable sources \cite{[3]}, which clearly indicates the need for new sustainability policies which will solely focus on increasing renewable source usage and decreasing fossil fuel production. This rapid economic growth highly affects the natural capital of the planet, by decreasing it and lowering the possibility of maintaining a healthy environment. 

The latest review of world energy \cite{[3]} indicates the growth in primary energy consumption in all world continents, with the highest consumption values generated in 2018 in Asia Pacific area (5985.8 mtoe), followed by North America (2832 mtoe), and Europe (2050.7 mtoe). In terms of the fuel type, North America, South America, Europe and Africa mostly use energy from oil sources, while the CIS (The Commonwealth of Independent States) and Middle East countries focus on using primary energy from natural gas \cite{[3]}. In Asia, a prevalent use of energy from coal sources is observed \cite{[3]}.   

In order to maintain the life-supporting systems of the planet, the sustainability concept has been introduced. Sustainability involves the ``development that meets the needs of the present without compromising the ability of future generations to meet their own needs" \cite{[4]}, and involves three dimensions: (1) economical, (2) social, and (3) environmental dimension, which are completely interdependent \cite{[5]}. If properly applied by creating sustainable policies and laws, the sustainability concept should preserve the Earth's ecosystems and allow optimized natural resource allocation. Unfortunately, human activities today exhaust the nature faster than it can regenerate, hence the ecological deficit arises as a critical problem when dealing with the climate change. As humans highly depend on nature, it is clear that such deficit should be handled in the near time future. 

There is a clear need to use nature's resources in accordance to their regenerative capacity, and to dispose waste in accordance to the speed of its absorption \cite{[6]}. But in order to do so, the availability of nature's resources and human requirements for natural resources should be estimated, hence the Ecological Footprint (EF) term has been introduced. The EF has been developed by Rees and Wackernagel \cite{[7]}, \cite{[8]} and presents ``the tool that enables us to estimate the resource consumption and waste assimilation requirements of a defined human population or economy in terms of a corresponding productive land area" \cite{[6]}.  It can also be interpreted as the demand of the population on the nature \cite{[9]}. In these terms, another concept is very closely connected to the EF and must be observed and analyzed in order to fully understand the state of natural resources and biosphere. Biocapacity shows the amount of biologically productive land and water area which can be used to provide humanity \cite{[9]}. Both measures are expressed in global hectares (gha) and are comparable. 

The EF and biocapacity in their calculations include several land use types, in particular: (1) cropland, (2) graying land, (3) fishing grounds, (4) forests, (5) uptake land, and (6) built-up land \cite{[9]}. Cropland includes the land area required to produce the agricultural products such as crop, livestock feeds, fish meals, oil crops and rubber \cite{[9]}.  Grazing land involves the land area covered in grass, i.e. grassland, used to provide foods for livestock, in addition to feeds from cropland area \cite{[9]}.  Fishing grounds involve fishery and aquaculture, while forest land involves the area covered in forests and used for harvesting fuelwoods and timber in order to supply forest products \cite{[9]}.  Carbon uptake land refers to the land used for carbon dioxide (CO$_2$) emissions absorption. As most of carbon uptake happens in forests, this is a subcategory of forest land \cite{[9]}.  Lastly, the built-up land involves the area covered in infrastructure such as housing, transportation, industrial buildings and reservoirs for hydroelectric power generation \cite{[9]}. Furthermore, there are two types of the EF: (1) the EF of production, and (2) the EF of consumption \cite{[10]}. The first presents the primary demand for natural resources \cite{[9]}, while the latter includes the area used to produce products to support an observed population's consumption habits \cite{[10]}. 

\begin{figure}[ht]
\begin{center}
\includegraphics[height=7.5cm, width=7.5cm, keepaspectratio]{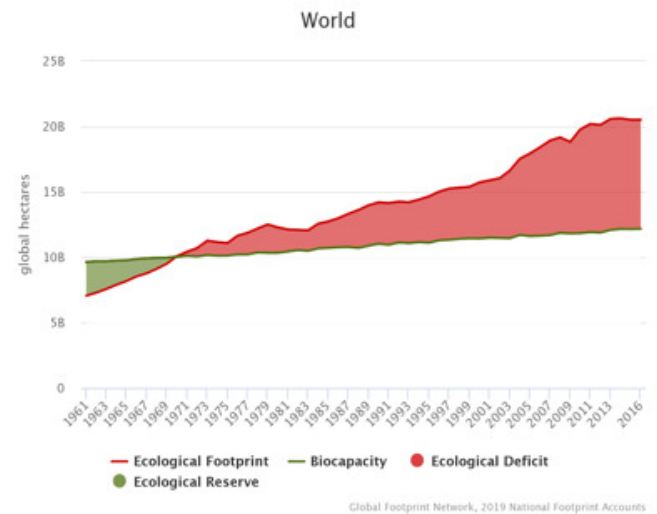}
\caption{EF and Biocapacity (source: \cite{[11]})}
\label{fig1}
\end{center}
\end{figure}

The importance of the EF is reflected in the fact that it mainly focuses on the needs of the planet and provides clear approximation of the impact of human demand for natural resources. Natural capital is very much limited and should be carefully and efficiently used, but not abused. Based on the last available data \cite{[11]}, there is a global biocapacity deficit, while EF is on the rise (Fig. \ref{fig1}). It can be observed that there is a very small rising trend of ecological reserves, but in order to achieve bigger increases, the pressure human activities put on the nature should weaken and allow nature to regenerate.

In \cite{[12]} the EF of Slovenia was investigated, and the values of the EF were predicted based on the energy consumption growth in terms of fossil fuels, electric energy, biomass, import and export of electrical energy and embodied energy in exported products. Energy consumption growth was calculated by predicting the ratio of industrial growth which was then used as a weighting factor for energy forecasting \cite{[12]}. The predictions indicated an increase in energy consumption in Slovenia, followed by the increase of the EF \cite{[12]}. In \cite{[13]} an urban EF prediction model based on the Markov chain was developed. Some of the indicators used in developing the model involve population, urbanization rate, total energy consumption, consumer price index, contribution share of third strata industries, and contribution share of total retail sales of consumer goods to the gross domestic product \cite{[13]}. It was shown that the energy consumption widely contributes the urban EF \cite{[13]}. In \cite{[14]} the environmental Kuznets curve for EF was investigated from the aspect of energy and financial development. It was found that high levels of energy use generate higher values of the EF and increase the environmental degradation \cite{[14]}.

The novelty of this approach is presented in the fact that the forecasting of the EF is based specifically and only on energy parameters, in particular on the primary energy consumption values. The developed prediction model gives some estimates in terms of future values of observed variables and can help policy makers to create efficient and sustainable decisions. Moreover, the model makes predictions based on the values of all variables, hence the relations between the variables are also considered when forecasting. 

This paper is organized as follows. Section \ref{sec2} presents the data and methodology used in this research. Section \ref{sec3} shows the obtained results and predictions of the developed model. Lastly, Section \ref{sec4} presents the conclusions of this research.

\section{Methodology}\label{sec2}
The dataset used in this research was obtained from several sources \cite{[11]},\cite{[15]},\cite{[16]} and involves global yearly data for the period 1971-2014. One dependent variable and eight independent variables were analyzed. The dependent variable represents the total EF of consumption, while the independent variables represent the primary energy consumption from different sources, in particular: natural gas, coal, oil, nuclear, hydroelectric, wind, solar photovoltaic (PV), and other renewable sources. The dependent variable, the total EF of consumption, is expressed in gha, while all independent variables are expressed in terawatt-hours. All analyses and modellings were performed using R programming language for statistical computing on a 1.7GHz dual-core Windows machine. 

The first step, after importing the variables, was to identify correlations between the dependent and independent variables. Correlations were calculated based on the Pearson's correlation coefficient (denoted $r$) which shows the strength of the relationship between the variables, if such exists \cite{[17]}. The values of the Pearson's correlation can be in the range [-1,1], where the value of 1 represents the strong, linear, relationship, while 0 shows no linear correlation. Depending on the sign, correlation can be positive or negative. 

The next step included data preprocessing, where each variable was separately log transformed. Because the primary energy consumption from solar PV sources, and from wind sources, included zero values, the log10 transformation was used to transform these variables, with added constant of 1. Other variables included values higher than 0, hence the standard log10 transformation was used in order to prepare the variables for time series prediction. After fitting the model and generating predictions, an inverse log transformation was performed in order to generate real-value predictions. 

The next step involved creating and fitting the Vector Autoregression model (VAR). The optimal number of lags was estimated based on the Akaike Information Criterion (AIC), and a value of 2 lags was used for creating the model. The AIC technique estimates the quality of a model when predicting the future values in terms that it compares each model to the other models. The AIC is calculated as follows \cite{[18]}:
\begin{equation}
AIC=-2logL+2k, 	
\end{equation}
where $L$ is the maximum likelihood value, and $k$ is the number of estimated parameters \cite{[18]}. 

The VAR is a linear model consisting of $n$-variables, in which every variable is explained by its own lagged values, and the values of other variables \cite{[6]}. Therefore, VAR is used to model interdependencies between multiple time series, hence it represents a multivariate model. A $p$-lag VAR can be represented as follows \cite{[19]}, \cite{[20]}:
\begin{equation}
y_t=c+A_{1y_{t-1}}+A_{2y_{t-2}}+...+A_{py_{t-p}}+e_t,
\end{equation}
where $c$ represents a vector of constants, $A_i$ represents the coefficient matrix, $y_{t-i}$ is the $i$-th lag of $y$, and $e_t$ is the vector of errors.

After fitting the model, predictions were generated for the period 2015-2024.

\section{Results and Discussion}\label{sec3}
Before performing the time series vector autoregression, a correlation analysis was executed. The obtained results are presented in Table \ref{tab1}, where it can be observed that the variables total EF of consumption and primary energy consumption from gas sources have the highest, positive correlation, with a value of $r$=0.995, followed by the correlation value obtained for the total EF of consumption and coal sources ($r$=0.986). The same value of correlation is also observed between the total EF of consumption and primary energy consumption from hydroelectric sources ($r$=0.986). For all other variables, correlation coefficients are also found to be high, except for the primary energy consumption from Solar Photovoltaic sources ($r$=0.621). All correlations are found to be significant ($p<$0.001). 
\begin{table*}[t!]
\caption{Correlation between the variables}
\begin{center}
\includegraphics[height=13cm, width=13cm, keepaspectratio]{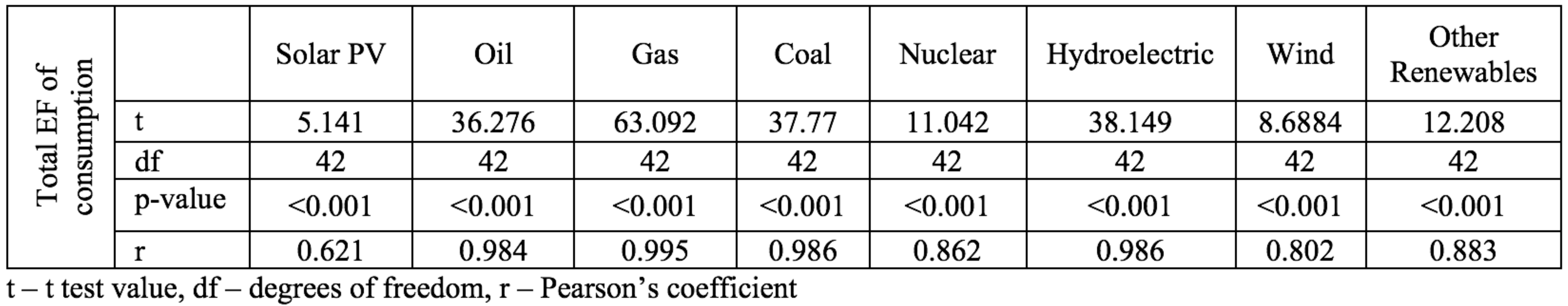}
\end{center}
\label{tab1}
\end{table*}%

\begin{figure*}[t!]
\begin{center}
\subfigure{\includegraphics[height=7cm, width=7cm, keepaspectratio]{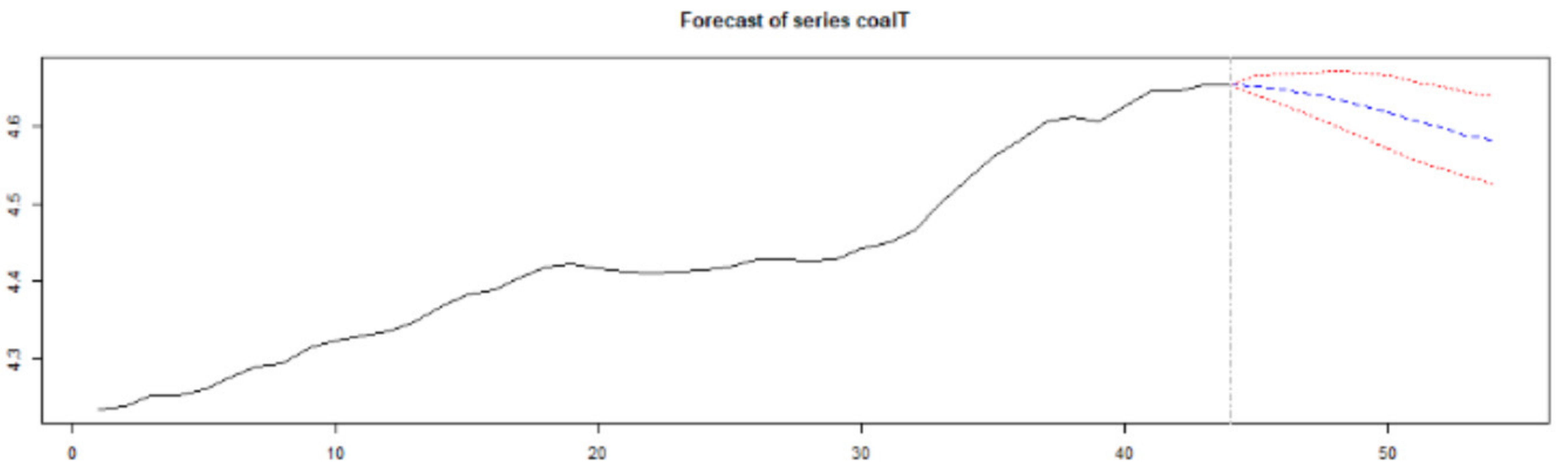}}\\
\subfigure{\includegraphics[height=7cm, width=7cm, keepaspectratio]{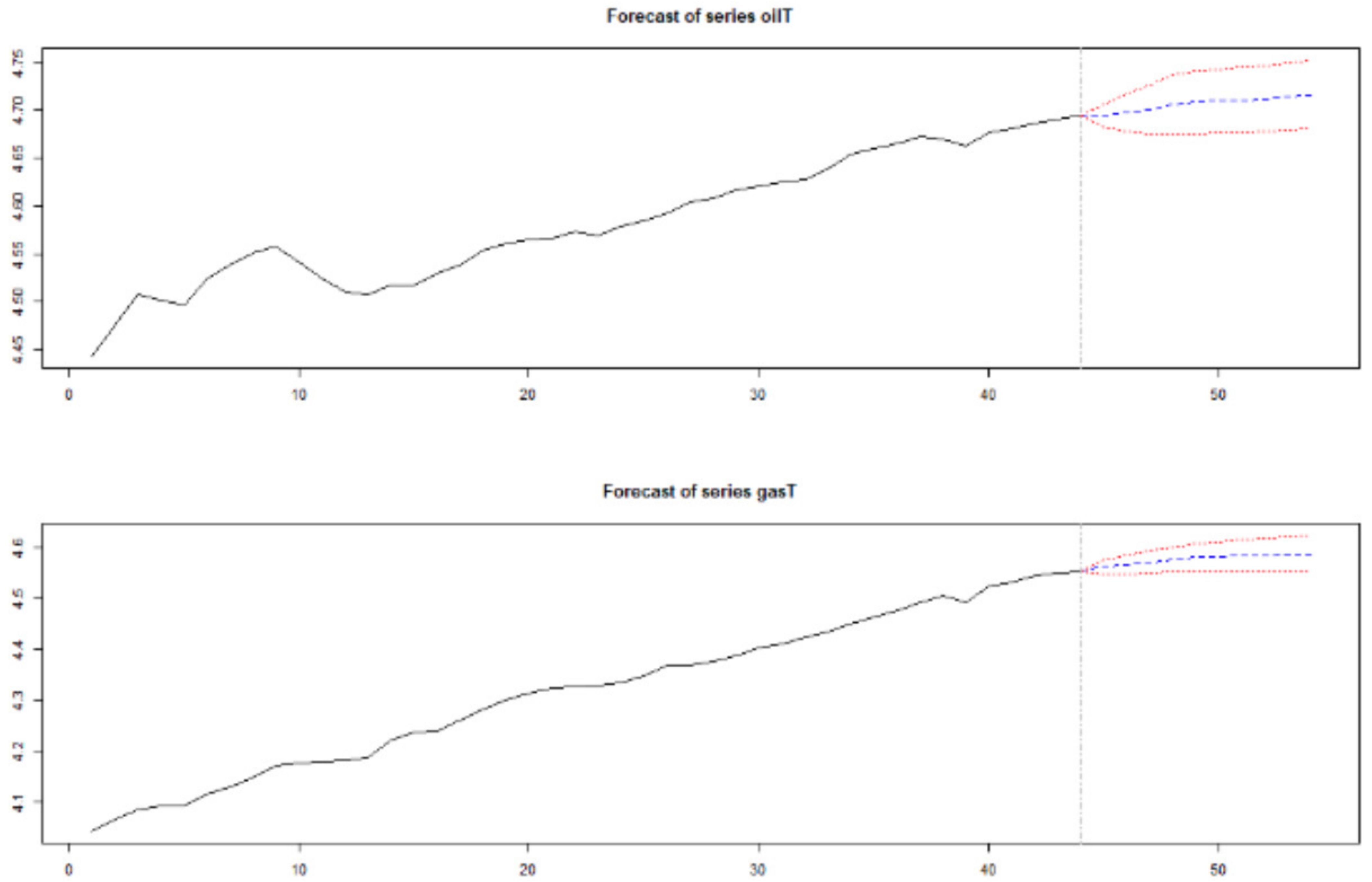}}
\subfigure{\includegraphics[height=7cm, width=7cm, keepaspectratio]{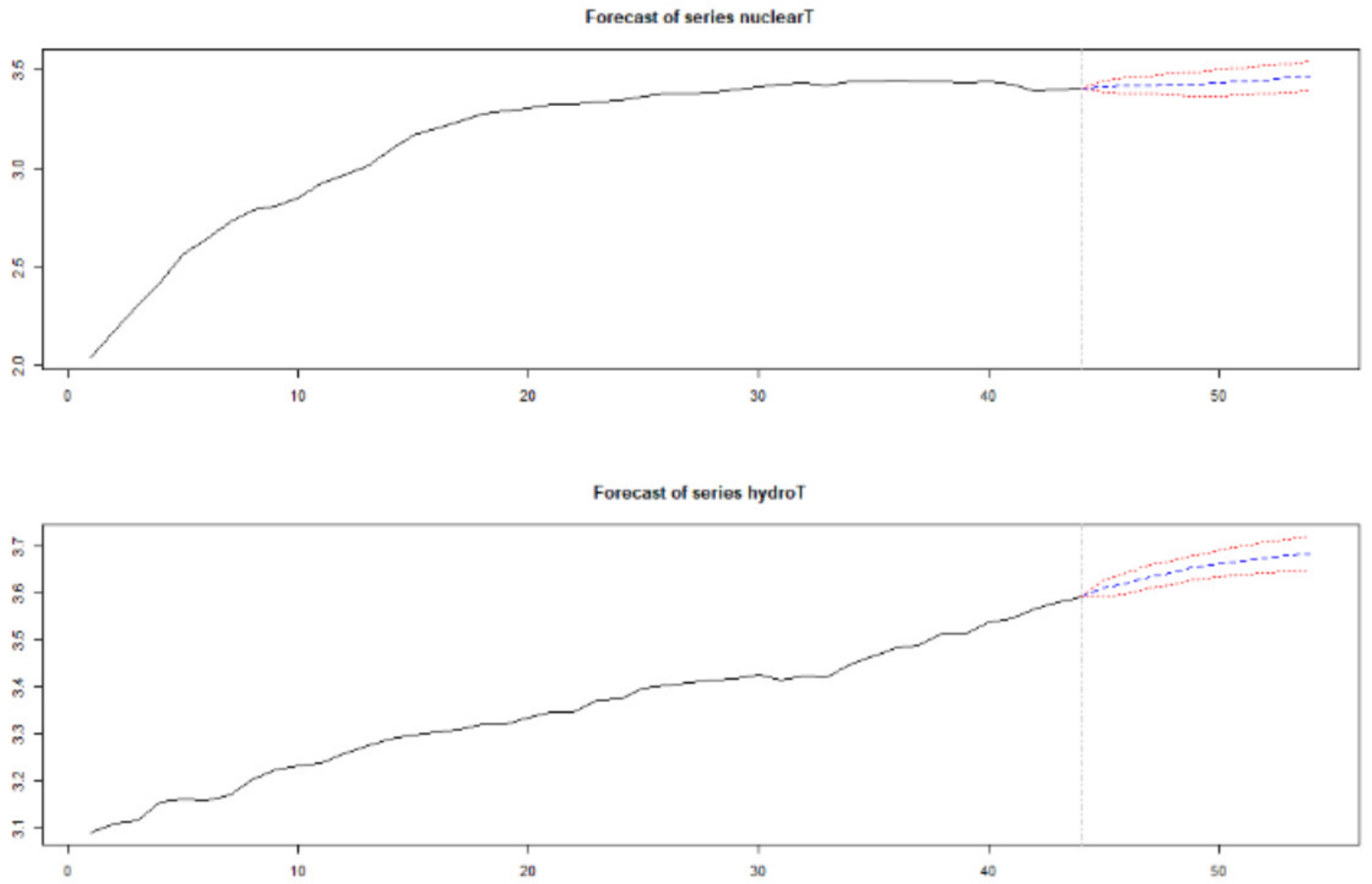}}\\
\subfigure{\includegraphics[height=7cm, width=7cm, keepaspectratio]{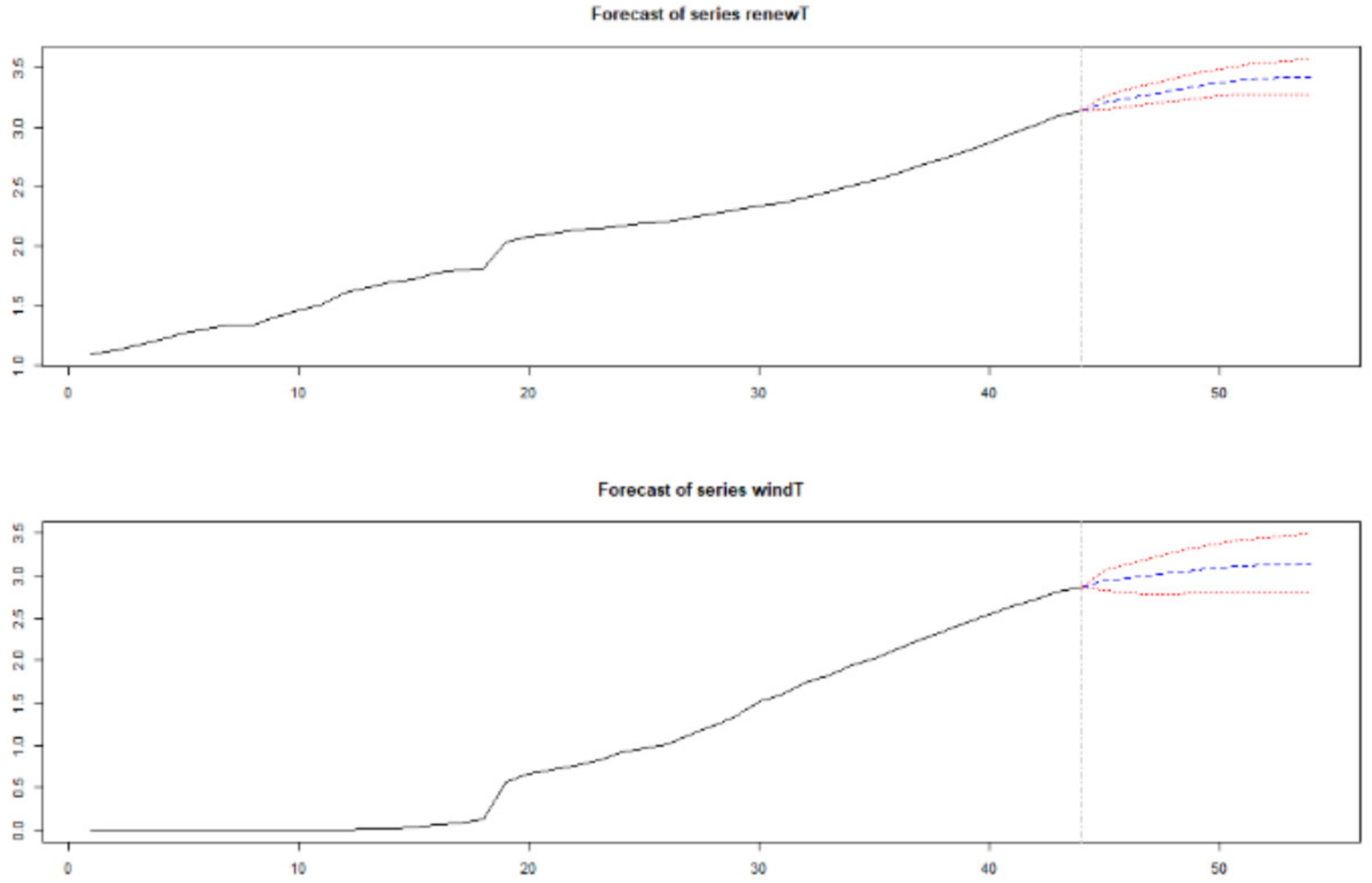}}
\subfigure{\includegraphics[height=7cm, width=7cm, keepaspectratio]{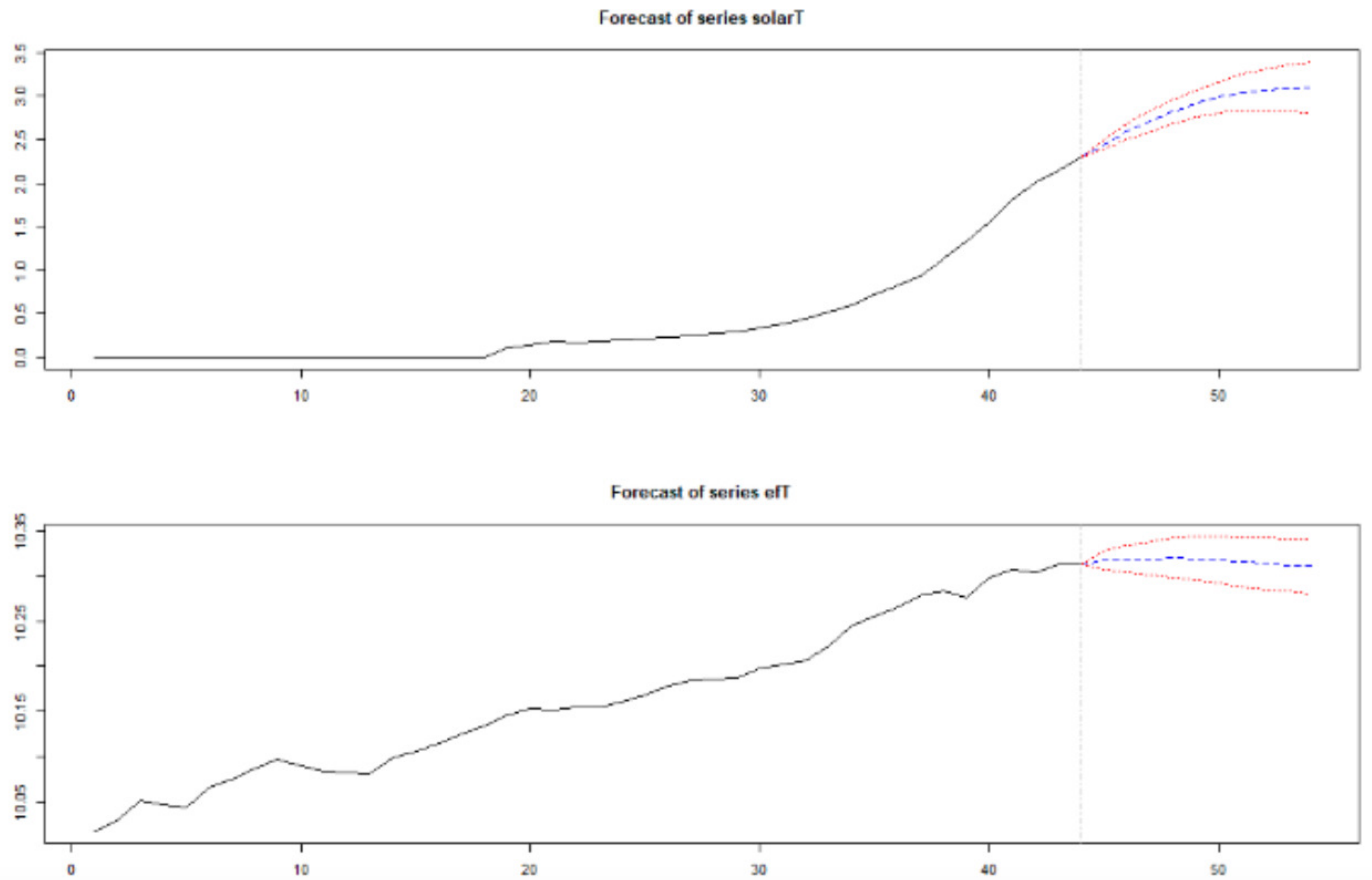}}
\caption{VAR forecast for each variable (log transformed values for coal, oil, nuclear, gas, hydro, other renewables, solar, wind, EF)}
\label{fig2}
\end{center}
\end{figure*}

\begin{figure*}
\begin{center}
\subfigure{\includegraphics[height=10cm, width=10cm, keepaspectratio]{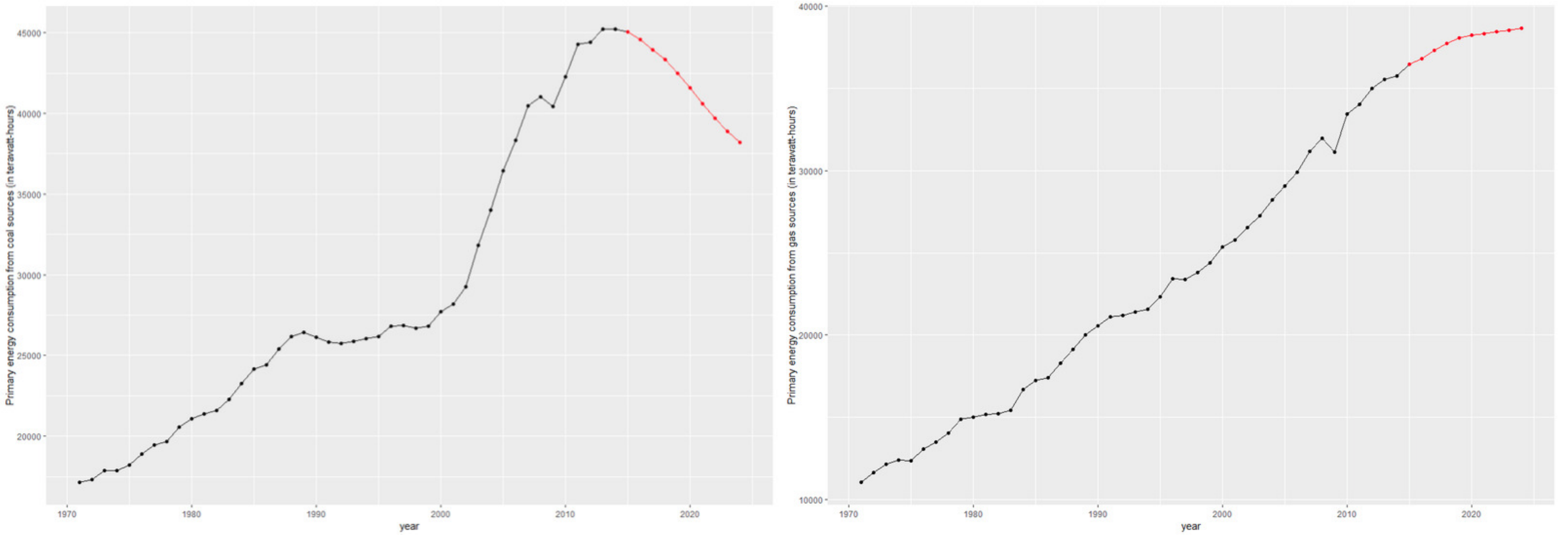}}
\subfigure{\includegraphics[height=10cm, width=10cm, keepaspectratio]{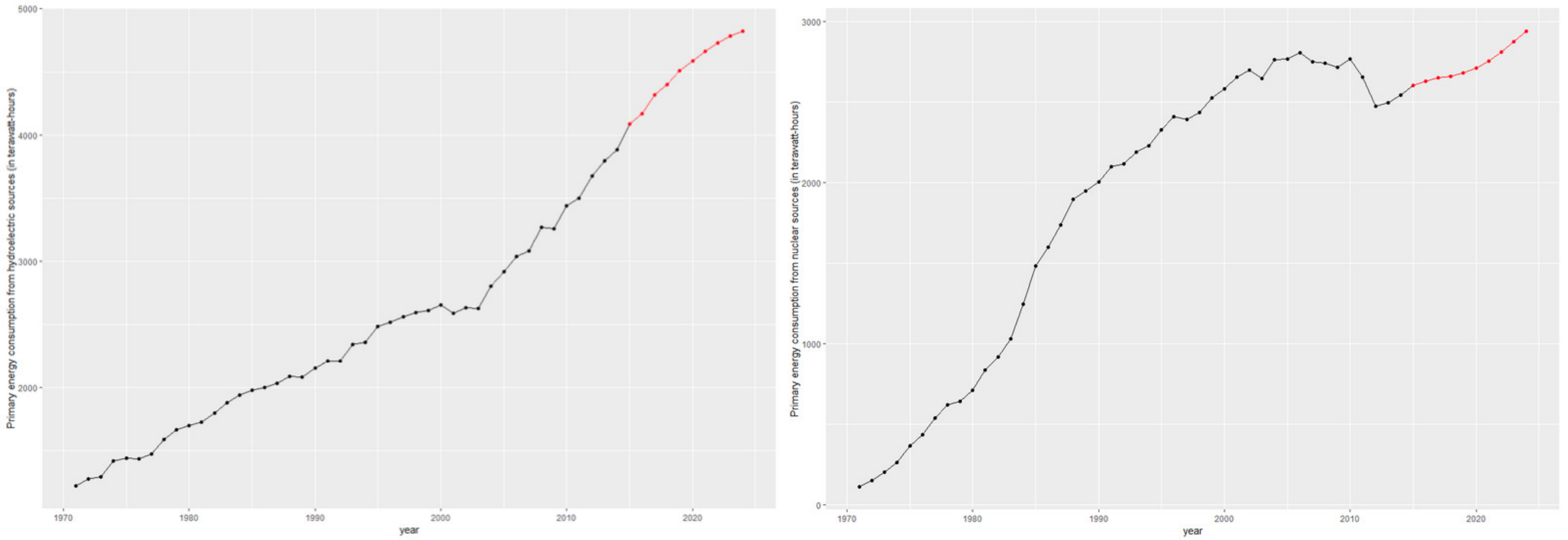}}
\subfigure{\includegraphics[height=10cm, width=10cm, keepaspectratio]{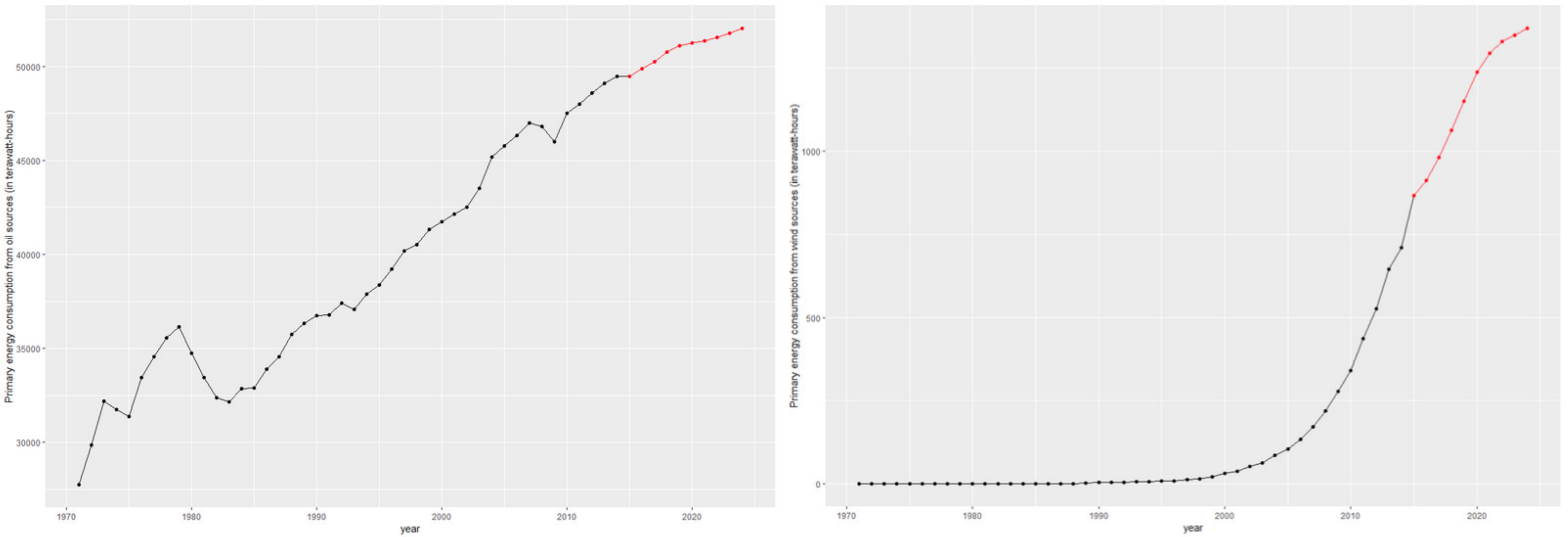}}
\subfigure{\includegraphics[height=10cm, width=10cm, keepaspectratio]{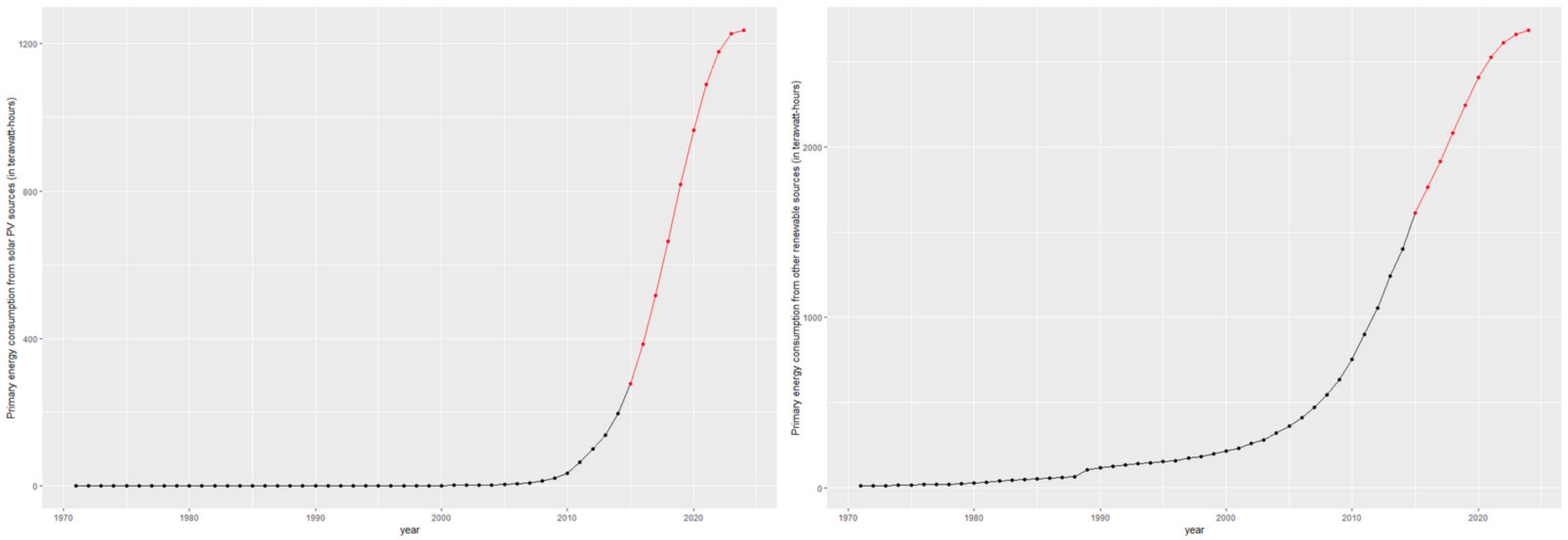}}\\
\subfigure{\includegraphics[height=5cm, width=5cm, keepaspectratio]{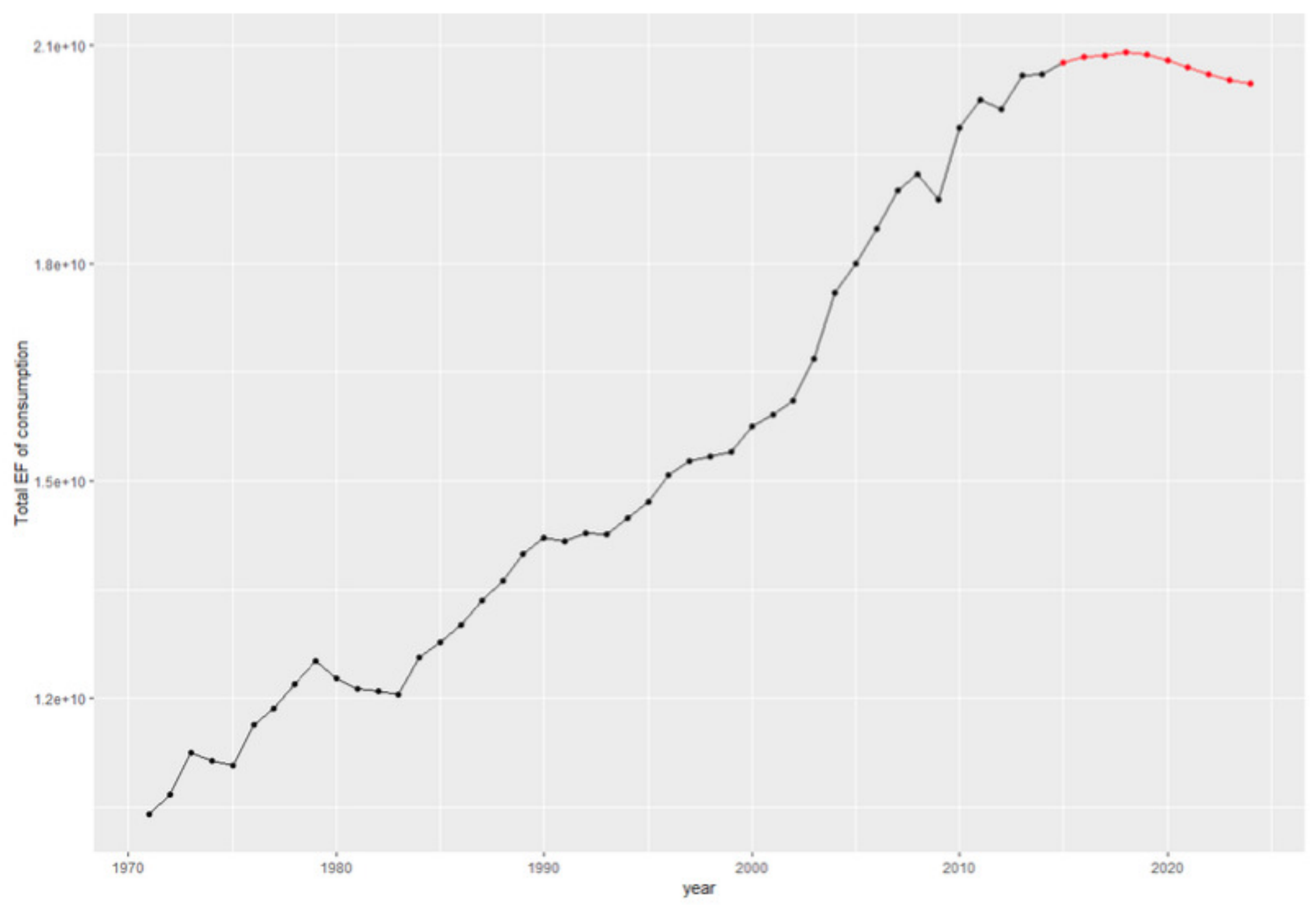}}
\caption{Real value predictions generated by the VAR model for each variable}
\label{fig3}
\end{center}
\end{figure*}

After log-transforming each variable, the data was fitted to the VAR model, and the graphs showing the real values, upper and lower prediction limits, and forecasted values were generated (Figure \ref{fig2}). Lastly, the predicted values were inverse log-transformed, and the results are shown in Figure \ref{fig3}. For each variable, the prediction trend is shown in Fig. \ref{fig3}.  

The predictions in Figure \ref{fig3} indicate that the amount of primary energy consumption from gas, hydroelectric, oil, nuclear, and wind sources will have an increasing trend in the upcoming years. These predictions support previous findings by other authors, such as the use of oil as an energy source which is predicted to increase in the future \cite{[21]}, \cite{[22]}, \cite{[23]}. Moreover, from Figure \ref{fig3} it is observable that the amount of primary energy consumption from coal sources will decrease in the following years, supporting the findings in \cite{[21]}, \cite{[22]}, \cite{[23]}. The use of energy from solar PV sources, and from other renewable sources will increase by the year of 2023, after which it will have a more stable trend. Lastly, considering the total EF of consumption is predicted to increase until 2019, when a slow decrease occurs. 

The obtained findings indicate a strong relationship between the energy and the ecological footprint, as with the decrease of fossil fuel consumption, the EF also decreases. 
\section{Conclusions}\label{sec4}
This paper analyzed the relationship between the total EF of consumption and the primary energy consumption from coal, hydroelectric, nuclear, oil, natural gas, wind, solar PV, and other renewable sources. The objective of the paper was to create a forecasting model for the EF prediction based only on energy parameters. The prediction model was developed using vector autoregression, where the values of each of the variables were predicted for the period 2015-2024. 

The results showed a high correlation between the EF of consumption and the primary energy consumption from all sources. Moreover, the prediction model suggests that the global EF will maintain an increasing trend until 2020, when it slowly starts to decline. Energy sources will maintain a high level of use in the future, with exception of the coal. Based on the predictions, global primary energy consumption from coal sources will decrease until 2024. As coal produces the highest concentrations of carbon dioxide, such predictions are positive. 

As more countries set sustainable goals and work towards accomplishing them, further action is still needed. Natural capital should be appropriated on a rational level, considering its boundaries. In terms of energy, the carbon footprint should be of special importance as it considers the effects of each phase of the life cycle of fuel on the environment.

\begin{center}
{\bf Acknowledgement}
\end{center}
This research was financially supported by the Mathematical Institute of the Serbian Academy of Sciences and Arts (Project III44006) and through the project of the Ministry of Education, Science and Technological Development of Serbia --TR34023.

\end{document}